\documentclass{JINST}

\newcommand{\hetrois}    {\mbox{$ ^{3}{\mathrm{He}}                            $}}

\newcommand{\neutt}{$\tilde{\chi}$}

\title{A prototype of a directional detector for non-baryonic dark matter search: MIMAC (~Micro-TPC Matrix of Chambers)}

\author{C. Grignon$^a$\thanks{Corresponding author.}~, G. Bernard$^a$, J. Billard$^a$, G. Bosson$^a$, O. Bourrion$^a$, O. Guillaudin$^a$, C.~Koumeir$^a$, F. Mayet$^a$, D.~ Santos$^a$,
 P. Colas$^b$,  E. Ferrer$^b$, I. Giomataris$^b$, A.~Allaoua$^c$, L. Lebreton$^c$  \\
\llap{$^a$}LPSC, Universit\'e Joseph Fourier Grenoble 1,CNRS/IN2P3, Institut Polytechnique de Grenoble,\\
  53 avenue des Martyrs, 38026 Grenoble, France\\
\llap{$^b$}IRFU/DSM/CEA,\\
  CEA Saclay, 91191 Gif-sur-Yvette cedex\\
\llap{$^c$}IRSN\\
  13115 Saint-Paul-Lez-Durance \\
  E-mail: \email{cyril.grignon@lpsc.in2p3.fr}}

\abstract{We have developed a micro-tpc using a pixelized bulk micromegas coupled to dedicated acquisition electronics as a read-out allowing 
to reconstruct the three dimensional track of a few keV recoils. The prototype has been tested with the Amande facility at the IRSN-Cadarache providing 
monochromatic neutrons. The first results concerning discrimination of a few keV electrons and proton recoils are presented.}


\begin{document}

\section{Dark matter directional detection}

Many astrophysical observations lead to the fact that dark matter represents an important part of the content of the universe.
These piece of evidence are present at different scales: locally, from the rotation curves of spiral galaxies \cite{rubin} 
or the Bullet cluster \cite{clowe} and on the largest scales, from  cosmological  observations \cite{wmap,archeops}. 
Most of the matter in the Universe consists of cold non-baryonic dark matter (CDM), the 
leading candidate for this class of yet undiscovered particles (WIMPs)  being the lightest supersymmetric particle. 
In various supersymmetric scenarii (SUSY), this neutral and colorless particle is the lightest neutralino \neutt.

In order to detect this particle, tremendous experimental efforts with a host of techniques have been made.
However, the main challenge for dark matter detection experiments remains present: discriminate the genuine WIMP signal from various background events such 
as $\alpha$, $\beta$, $\gamma$, $\mu$ and neutrons.

To solve this problem and to demonstrate that the observed recoils are indeed due to WIMPs, direct detection experiments need a clear signal that can not 
be mimic by background. 
The most promising strategy is to to search for a favored incoming direction for the WIMP signal, this signature beeing the only one that allows to clearly 
correlate the recoil events to the galactic halo of WIMPs.
Indeed, our motion with respect to the Galactic rest frame produces a direction dependance in the recoil spectrum, the WIMP flux should come from the 
direction of the solar motion, towards the Cygnus constellation \cite{directionality,agreen}. 
Several projects aiming at directional detection  of Dark Matter are being developped \cite{Drift,mit,MIMAC}. 


Gaseous $\mu$TPC detectors present the privileged features of being able to reconstruct the track of the recoil following the interaction, thus allowing 
to access both the energy and the track properties (lenght and direction). 
In order to reconstruct precisely both low energy (few keV) and the 3D track of a nuclear recoil (few mm), we chose to use a micro-pattern gaseous detector,
and more precisely a bulk micromegas \cite{bulk}.

\section{The MIMAC project and the $\mu$TPC}

The MIMAC project is based on a matrix of gaseous $\mu$TPC, filled either with \hetrois~, $\rm CF_4$, $\rm CH_4$ or $\rm C_4H_{10}$.
Working at a low pressure regime will allow us to discriminate electrons from nuclei recoils by using both energy and track length information, an 
electron track being longer than the one of a nucleus for a given energy.
Besides, identification of neutrons will be done by performing a time correlation of chambers, assuming that a WIMP will not interact twice
in the whole MIMAC detector.
Ultimately, the last background rejection tool is the reconstruction of the incoming direction of the particle.

Finally, we need to precisely measure energy of a few keV and to reconstruct 3D tracks of a few mm.
To do so, we developped a $\mu$TPC prototype with a 16.5 cm drift space associated to a bulk micromegas responsible for the detection of electrons.
At first, we used a standard 128 $\mu$m bulk micromegas (non-pixelized anode plane) in order to measure the energy resolution of our detector.

To correctely assess the real recoil energy of the nucleus, we performed a complete measurement of the ionization quenching factor in the energy range of 
dark matter search, i.e. below 10 keV. As described in \cite{prl},\cite{ogparis},\cite{fmparis}, we designed an Electron Cyclotron Resonance Ion Source (ECRIS), 
able to produced various ions (proton, $^3$He, $^4$He, $^{19}$F) with an energy from 1 keV to 50 keV.
We successfully measured the IQF in helium at low energies, and reached an energy threshold below 1 keV, which is a key point for Dark Matter detectors, 
since it is needed to evaluate the nucleus recoil energy and hence the WIMP kinematics.

Although a precise  measurement of the energy of the recoil is the starting point of any background discrimination, the 3D reconstruction of the track
is necessary to detect the direction of dark matter particle.

\section{The pixelized bulk micromegas and the MIMAC acquisition electronics}

In order to reconstruct a few mm track in three dimensions, we decided to use a bulk micromegas \cite{Giomataris95,bulk}  with a 3 cm by 3 cm active area, 
segmented in 300 $\mu$m pixels.
We use a 2D readout with 424 $\mu$m pitch, in order to read both dimensions (X and Y).
This bulk is provided with a 325 LPI (Line Per Inch) weaved 25 $\mu$m thick stainless steel micro-mesh.

In order to reconstruct the recoil of a nucleus in 3D, we needed a dedicated data acquisition electronics, self-triggered and able to perform the anode 
sampling at a frequency of 40 MHz \cite{richer}.
To do so, the LPSC electronics team designed from scratch an ASIC in a 0.35 $\mu$m BiCMOS-SiGe technology.
With an area approximatively equal to 15 mm$^2$, this ASIC contains 16 channels, each of these channels having its own charge-sensitive preamplifier, 
current comparator and 5 bit coded tunable threshold.
The 16 channels are sent to a mixer and a shaper to measure the energy in the ASIC.
Each of the 12 ASICs is connected to FPGAs programmed to process, merge and time sort data. 
Finally, the electronic board is connected to an Ethernet microcontroller which forwards the data via a TCP socket server to the acquisition station.
This first version of the MIMAC ASIC is up and running in the MIMAC prototype since May 2008 and the next version, with 64 channels, is currently under 
development at the LPSC and will be ready before the end of 2009.

\section{3D track recontruction}
   
\begin{figure}[h!]
\begin{center}
\includegraphics[scale=0.65]{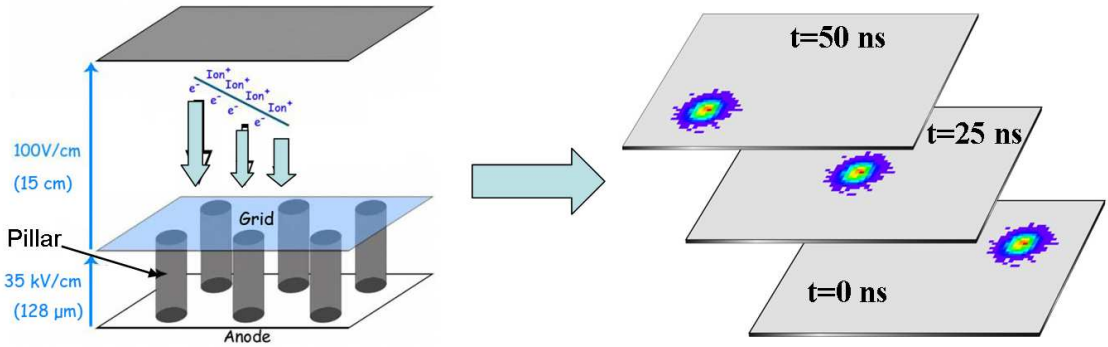}
\caption{Track reconstruction in MIMAC. The anode is scanned every 25 ns and the
3D track is recontructed, knowing the drift velocity,  from the series of images
of the anode.}
\label{recon}
\end{center}
\end{figure}

One key point of a dark matter directional detector project is to show the possibility to reconstruct a 3D track, as 
the required exposure to sign any directional asymmetry is decreased by an order of magnitude between  2D read-out and 3D read-out \cite{agreen}.
With this dedicated electronics associated to the pixelized bulk micromegas, we are able to reconstruct the recoil of the nucleus in 3D.
As pictured on figure  \ref{recon}, the electrons move towards the grid in the drift space and are projected on the anode thus allowing to access 
information on x and y coordinates.
The third coordinate is obtained by sampling the anode every 25 ns and by using the knowledge of the drift velocity of the electrons.

We developped a simulation software to test both the capabilities of the DAQ to reconstruct tracks and the reconstruction algorithm itself.
For a realistic detector instrumented with our electronics, we show that the 3D track reconstruction can be achieved with a rather good resolution, 
of $\rm 0.3 mm$ on the track length and below $\rm 4^\circ$ on $\theta$ and $\phi$, assuming a straight trajectory for the recoil ion \cite{simucyril}.
The first experimental data obtained with our prototype are quite promising.

\begin{figure}[h!]
\begin{center}
 \includegraphics[scale=0.4]{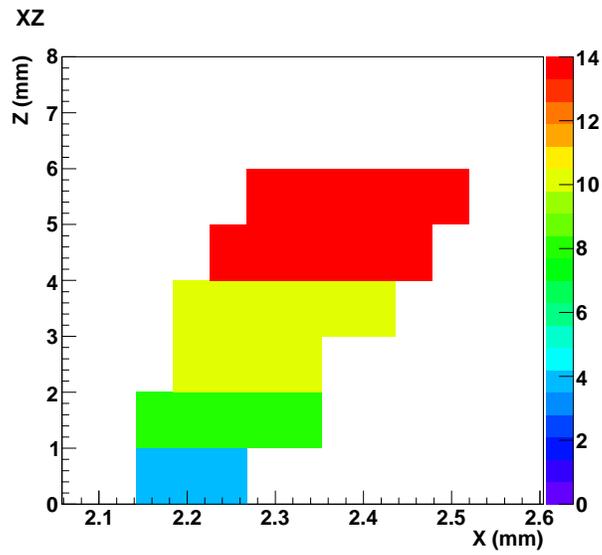}
 \caption{Reconstructed track of a proton recoil created by a 144 keV neutron in 100 mbar $C_4H_{10}$ projected on the XZ plane}
 \label{tracks}
\end{center}
\end{figure}

As one can see on figure \ref{tracks}, we were able to reconstruct nuclei recoils created by 144 keV neutrons at the Amande facility (IRSN-Cadarache) in
pure isobutane at 100 mbar.

\begin{figure}[h!]
\begin{center}
 \includegraphics[scale=0.7]{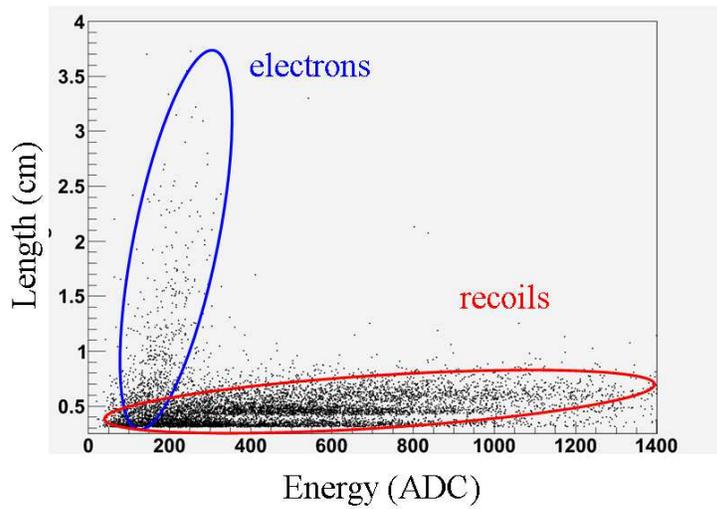}
 \caption{Length versus energy plot for electrons and neutrons in pure isobutane at 100 mbar}
 \label{plot}
\end{center}
\end{figure}

Moreover, on the length/energy plot (see figure \ref{plot}), we can see a clear separation of the electron noise from the nuclei recoils signal.  

\section{Conclusion}

We were able to precisely measure the recoil energy of an helium nucleus down to 1 keV, thanks to ECR ion source developped at the LPSC.
We developped a complete dedicated electronic system (ASIC, FPGA, Ethernet card and DAQ system) in order to perform a 40 MHz sampling of the pixelized bulk
micromegas and to finally reconstruct a 3D track of the recoil nucleus.
We reconstructed tracks of 5.9 keV electrons, as well as proton recoils issued from 144 keV neutron elastic interactions.
Discrimination of recoils from electron background is possible.

\end{document}